\documentclass[conference,10pt]{IEEEtran}
\IEEEoverridecommandlockouts
 
\usepackage{graphicx}
\usepackage{pgfplots}
\usepackage{pgfplotstable}
\usepackage{booktabs}
\usepackage{cite}
\usepackage{amsmath,amssymb,amsfonts}
\usepackage{algorithmic}
\usepackage{graphicx}
\usepackage{textcomp}
\usepackage{xcolor}
\def\BibTeX{{\rm B\kern-.05em{\sc i\kern-.025em b}\kern-.08em
    T\kern-.1667em\lower.7ex\hbox{E}\kern-.125emX}}
\begin{document}

\title{Multilingual Standalone Trustworthy Voice-Based Social Network for Disaster Situations
\vspace{3em} %
 
}

\author{\IEEEauthorblockN{1\textsuperscript{st} Majid Behravan}
\IEEEauthorblockA{\textit{Department of Computer Science} \\
\textit{Virginia Tech}\\
Blacksburg, VA \\
0000-0001-6525-6646}
\and

\IEEEauthorblockN{2\textsuperscript{nd}Elham Mohammadrezaei}
\IEEEauthorblockA{\textit{Department of Computer Science} \\
\textit{Virginia Tech}\\
Blacksburg, VA \\
0009-0002-8421-1165}
\and

\IEEEauthorblockN{3\textsuperscript{nd} Mohamed Azab}
\IEEEauthorblockA{\textit{Department of Computer and Information Sciences} \\
\textit{Virginia Military Institute}\\
VA \\
0000-0002-9386-4726}
\and

\IEEEauthorblockN{4\textsuperscript{th} Denis Gra{\v{c}}anin}
\IEEEauthorblockA{\textit{Department of Computer Science} \\
\textit{Virginia Tech}\\
Blacksburg, VA \\
0000-0001-6831-2818}}

\maketitle

\begin{abstract}

In disaster scenarios, effective communication is crucial, yet language barriers often hinder timely and accurate information dissemination, exacerbating vulnerabilities and complicating response efforts. This paper presents a novel, multilingual, voice-based social network specifically designed to address these challenges. The proposed system integrates advanced artificial intelligence (AI) with blockchain technology to enable secure, asynchronous voice communication across multiple languages. The application operates independently of external servers, ensuring reliability even in compromised environments by functioning offline through local networks. Key features include AI-driven real-time translation of voice messages, ensuring seamless cross-linguistic communication, and blockchain-enabled storage for secure, immutable records of all interactions, safeguarding message integrity. Designed for cross-platform use, the system offers consistent performance across devices, from mobile phones to desktops, making it highly adaptable in diverse disaster situations. Evaluation metrics demonstrate high accuracy in speech recognition and translation, low latency, and user satisfaction, validating the system's effectiveness in enhancing communication during crises. This solution represents a significant advancement in disaster communication, bridging language gaps to support more inclusive and efficient emergency response.
\end{abstract}


\section{Introduction}
In today's interconnected world, effective communication is essential, particularly during disaster situations where timely information dissemination can save lives. However, linguistic diversity often poses significant challenges in ensuring that crucial messages are understood by all stakeholders. Traditional communication platforms frequently fail to address these language barriers, leaving gaps in understanding that can have dire consequences during emergencies. The advent of AI technologies and decentralized systems like blockchain offers new possibilities for creating secure, multilingual communication networks that can function reliably even in the most challenging circumstances. Motivated by the need for a robust solution to bridge language gaps in crisis scenarios, this paper presents a novel, multi-platform application that integrates advanced AI with blockchain technology to enable trustworthy, asynchronous voice-based social media communication across language barriers. The application is designed to run seamlessly on various devices, including mobile phones, tablets, and desktops, and is compatible with different operating systems, ensuring accessibility and reliability in diverse environments.

The significance of addressing multilingual communication barriers in disaster situations cannot be overstated. Traditional approaches to disaster communication often fail to account for the linguistic diversity inherent in many societies, leading to what is termed "disaster linguicism." This phenomenon exacerbates the vulnerability of Indigenous, Tribal, Minority, and Minoritized peoples, as critical information is often not accessible in their native languages. By proposing a solution that integrates advanced AI with blockchain technology, this paper directly responds to the urgent need for more inclusive and effective disaster risk reduction strategies that consider the linguistic needs of all communities, thereby reducing social vulnerability and enhancing resilience during crises \cite{uekusa2023}.

Despite advancements in communication technology, there remains a critical gap in enabling asynchronous, multilingual voice communication during disaster situations. Current social media platforms and communication tools often rely on centralized servers and lack the necessary security features to ensure message integrity and authenticity, which are particularly vulnerable in disaster scenarios where external networks may fail. Additionally, the language translation tools available are not typically integrated into a system that provides both translation and secure message verification. This lack of integration can lead to misinformation, miscommunication, and delays in critical information sharing, which can exacerbate the effects of disasters. Furthermore, many existing solutions are platform-dependent, limiting their effectiveness across different devices and operating systems.

This paper presents significant advancements in the field of disaster communication and multilingual technology through the following key contributions:

\begin{itemize}
    \item \textbf{Innovative Application for Disaster Scenarios}: We introduce a novel, multi-platform application specifically designed to function reliably in disaster situations, where communication infrastructure may be compromised. The application’s ability to operate offline on a local network without requiring an internet connection ensures its deployment in critical environments. Its lightweight design allows it to run on portable devices like laptops, making it highly adaptable and infrastructure-free.

    \item \textbf{AI-Driven Asynchronous Multilingual Translation}: The application leverages advanced AI technology to facilitate asynchronous voice communication by automatically translating spoken messages across multiple languages. Furthermore, it synthesizes these translations back into voice, providing a natural and effective method of communication that transcends language barriers.

    \item \textbf{Blockchain-Enabled Secure Communication}: A key feature of this application is its implementation of secure storage and verification mechanisms using Ethereum blockchain. This ensures the integrity and authenticity of translated messages, critical in maintaining trustworthy communication during emergencies.

    \item \textbf{Cross-Platform Usability and Security}: The application is designed to operate independently of external servers, with all AI processing and blockchain management performed locally. This not only enhances security and reliability but also ensures a seamless user experience across various devices and operating systems, making it accessible in diverse disaster environments.

    \item \textbf{Robust and Portable Communication System}: By focusing on portability, trustworthiness, and transparency, this application provides a robust communication solution in emergency situations. It guarantees that communication remains secure, reliable, and effective, even when traditional systems fail.
\end{itemize}

\section{Related work}
\subsection{Multilingual Voice Social Media}
Multilingual voice social media is reshaping how individuals communicate across linguistic and cultural boundaries. These platforms enable seamless interactions through advanced speech recognition and real-time translation technologies. The challenge lies in ensuring that translations are not only accurate but also culturally nuanced, preserving the original intent and context. In this evolving landscape, multilingual voice platforms are essential for connecting diasporic communities and fostering meaningful dialogue in an increasingly globalized world\cite{jacquemet2019beyond}.

Examining existing platforms reveals a variety of tools that support these interactions. Voice-based social media platforms, such as Clubhouse and HearMeOut, provide users with audio-based interactions, offering a more personal communication style compared to traditional text-based media. Additionally, recording-based forums like Sangeet Swara and Baang cater to users with limited literacy and internet access, emphasizing inclusivity. These platforms facilitate authentic interactions by allowing users to express themselves through their voice, making online connections feel more genuine. However, reliance on recorded voice interactions poses challenges for those who prefer text-based content creation or have privacy concerns regarding voice recordings\cite{zhang2021social}.

Moreover, social media platforms have emerged as vital tools for grassroots collective action, enabling citizens to voice concerns and mobilize support. These platforms facilitate rapid information dissemination and community formation around shared causes. However, their effectiveness can be undermined by algorithmic biases and echo chambers, which often reinforce social divides rather than bridge them\cite{dumitrica2020voice}. The political value of voice highlights the necessity for technologies that can faithfully capture and translate speech nuances across different languages and dialects, ensuring inclusivity in digitally mediated collective actions. This underscores the importance of platforms that support both the articulation of voice and facilitate listening, acknowledging diverse perspectives\cite{dumitrica2020voice}. 

Despite existing solutions, they often lack robust mechanisms for verifying the authenticity of translated messages, leading to potential misinformation. Additionally, many platforms rely heavily on centralized systems, which may compromise security and reliability during disasters when external networks are unavailable. The integration of secure and decentralized technologies, such as blockchain, remains underexplored in enhancing the trustworthiness of multilingual interactions in social media.

\subsubsection{Speech Recognition and Translation Technologies}
Significant advancements in speech recognition and synthesis technologies, particularly with deep neural network models like WaveNet, have enhanced the effectiveness of voice assistants and smart devices. The primary challenge remains ensuring accurate recognition and translation across languages and dialects while preserving linguistic nuances. The design of the voice, whether human-like or robotic, also influences user interaction and trust, prompting researchers to focus on improving accuracy, user experience, and cultural sensitivity in voice interactions\cite{cambre2019one}.

The ability to capture and amplify voice—through spoken word in videos or written comments—plays a crucial role in these platforms as tools for activism. In the context of multilingual voice social media, technology must ensure that voices from diverse linguistic backgrounds are accurately represented and understood. This involves not just capturing speech but also translating it in a way that maintains the original intent and emotional impact, which is vital for sustaining engagement and participation\cite{mccosker2015social}.

User experience on platforms like YouTube is heavily influenced by features such as commenting, sharing, and liking, which impact content visibility. In a multilingual context, this complexity increases as users navigate linguistic and cultural differences. Multilingual voice social media platforms must ensure meaningful engagement across languages, requiring thoughtful content presentation, moderation, and interaction\cite{barnes2019social}. The challenges in ensuring that diverse voices are not only expressed but also understood are amplified by the need for accurate translation and the conveyance of cultural nuances. For voice to hold political value, platforms must prioritize both the expression and active engagement of these voices, fostering a more inclusive and democratic experience\cite{dumitrica2020voice}.

Despite advancements in speech recognition, existing systems often fall short in their ability to provide contextually accurate translations that reflect the emotional tone of the speaker. The need for integrating user feedback to refine translation algorithms is often overlooked, which can lead to dissatisfaction among users from diverse backgrounds.

\subsection{Multiplatform Mobile Applications}
The rapid proliferation of mobile devices and the increasing demand for seamless user experiences across different platforms have made multiplatform mobile applications a critical area of development. Users expect consistent performance and functionality on Android, iOS, or other platforms, which poses challenges for developers in creating applications that maintain uniformity while leveraging the unique capabilities of each platform\cite{delia2015multi}. This need has driven the adoption of cross-platform development frameworks, which allow for the efficient creation and maintenance of applications that function across multiple environments.

Cross-platform frameworks like React Native and Flutter enable developers to use a single codebase across multiple platforms, ensuring consistent usability, efficiency, and effectiveness. Usability testing methods, such as controlled observation and surveys, can further refine the user experience across devices\cite{weichbroth2020usability}. The architecture of these cross-platform systems is designed to optimize performance and responsiveness, which are vital for user satisfaction. Efficiency, a key usability attribute, emphasizes the need for mobile apps to load quickly and respond without delays. Developers must balance code reusability with platform-specific optimizations to enhance app responsiveness\cite{choi2009application}.

Performance and responsiveness are crucial in contexts where immediate interaction is expected, such as in e-commerce or customer service apps. Developers recommend using performance monitoring tools and continuous optimization to ensure that applications meet user expectations across all platforms\cite{weichbroth2020usability}. As mobile applications increasingly handle sensitive user data, the focus must shift to enhancing data security and leveraging blockchain technology to protect user privacy and ensure data integrity.

Most existing multiplatform applications struggle with providing a uniform user experience across devices due to differences in performance and capabilities. Furthermore, the integration of security features such as blockchain remains limited, exposing sensitive user data to potential breaches. The need for seamless and secure cross-platform communication is becoming increasingly critical in high-stakes environments.

\subsection{Data Security: Blockchain for Transcription and Translation}
Blockchain technology provides an innovative approach to ensuring data integrity, especially in environments where the authenticity and accuracy of data are critical. By utilizing a decentralized ledger, blockchain offers an immutable and transparent record of transactions, making unauthorized data modifications virtually impossible. For instance, the MedRec system demonstrates how blockchain can manage electronic medical records (EMRs) securely, ensuring that data across different platforms remains consistent and tamper-proof\cite{7573685}. Moreover, in sectors such as education and healthcare, blockchain has been applied to securely store and manage sensitive data, maintaining its integrity across various stages of processing and transmission\cite{yuda2023implementation}. 

By maintaining a verifiable audit trail, blockchain ensures that every transaction or modification is recorded, allowing for easy detection of discrepancies. This is particularly important in the context of multilingual transcription and translation, where maintaining the original meaning and accuracy across different languages is essential. By recording every step of the translation process on a blockchain, the integrity of both the original and translated content is preserved\cite{long2024study, 10480277}.

Blockchain significantly enhances privacy and security in multilingual applications. Its decentralized and cryptographic nature ensures that sensitive data, including multilingual transcripts and translations, is securely encrypted and accessible only to authorized users\cite{yuda2023implementation}. This capability is crucial in systems that handle personal information across multiple languages and cultural contexts.

While blockchain presents a strong solution for data integrity and security, its integration in real-time applications is still nascent. Many existing implementations do not fully leverage the capabilities of blockchain for secure, asynchronous communication in multilingual contexts. There is a need for systems that not only secure data but also ensure that data can be processed efficiently during high-demand situations, such as disasters.

\subsection{Local Server Deployment for Security and Emergency Mode}
In disaster management, local servers are vital for maintaining secure communication networks when traditional infrastructure fails. Operating independently, these servers ensure continuity and data integrity by integrating with ad hoc and satellite systems. They also offer significant security benefits, such as reduced exposure to external threats and better control of data flow, which is crucial during emergencies\cite{wang2023overview}. Additionally, activating an emergency mode on these servers enables quick deployment and resource scaling, ensuring critical communication and data management remain functional during urgent rescue efforts and real-time updates.

Local servers offer significant security advantages by reducing exposure to the broader internet and controlling access to sensitive data. In emergency mode, they quickly deploy essential services like real-time analytics and resource coordination, ensuring that critical operations remain effective during disasters\cite{ray2017internet}.

Damaševičius et al. utilized local servers to enhance real-time situational awareness and decision-making by hosting critical IoT devices and applications. This approach reduces latency and improves the speed and reliability of emergency responses, while also bolstering the security of sensitive information by keeping data within a closed, localized network, thereby minimizing the risk of cyber-attacks during vulnerable emergency situations\cite{damavsevivcius2023sensors}.

Despite the benefits of local servers, many disaster management systems still rely heavily on centralized cloud solutions, which can fail during crises. The integration of local servers with decentralized technologies like blockchain is often not explored, limiting their potential for enhancing security and reliability during emergencies. 

Significant advancements have been made in each of these areas, but integrating these technologies into a cohesive system requires careful consideration. The next section, System Design, will outline the architecture and components of the proposed system, showing how these technologies combine to meet specific needs\cite{cheimaras2023emergency}.

\section{System design and Implementation}
\begin{figure}[htbp]
    \centering
    \includegraphics[width=1\linewidth]{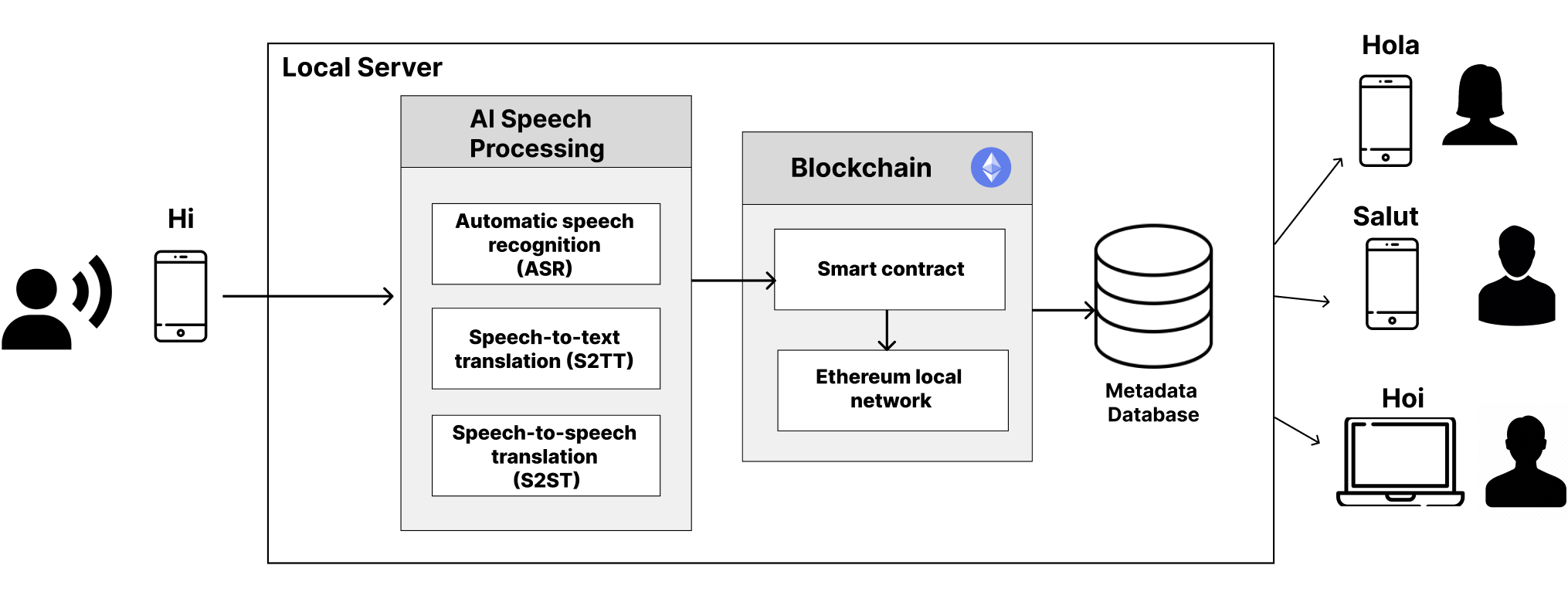}
    \caption{System Architecture for a Multilingual Voice-Based Social Network using Blockchain.}
    \label{fig:Framework}
\end{figure}

 The implementation comprises several key components and processes that enable seamless user interaction and message translation, all while ensuring data integrity and security via blockchain technology, as shown in Figure~\ref{fig:Framework}.

\subsection{User Registration and Profile Setup}

\begin{figure}[htbp]
    \centering
    \hspace*{.2in}\begin{minipage}{0.2\textwidth}
        \centering
        \includegraphics[width=\linewidth]{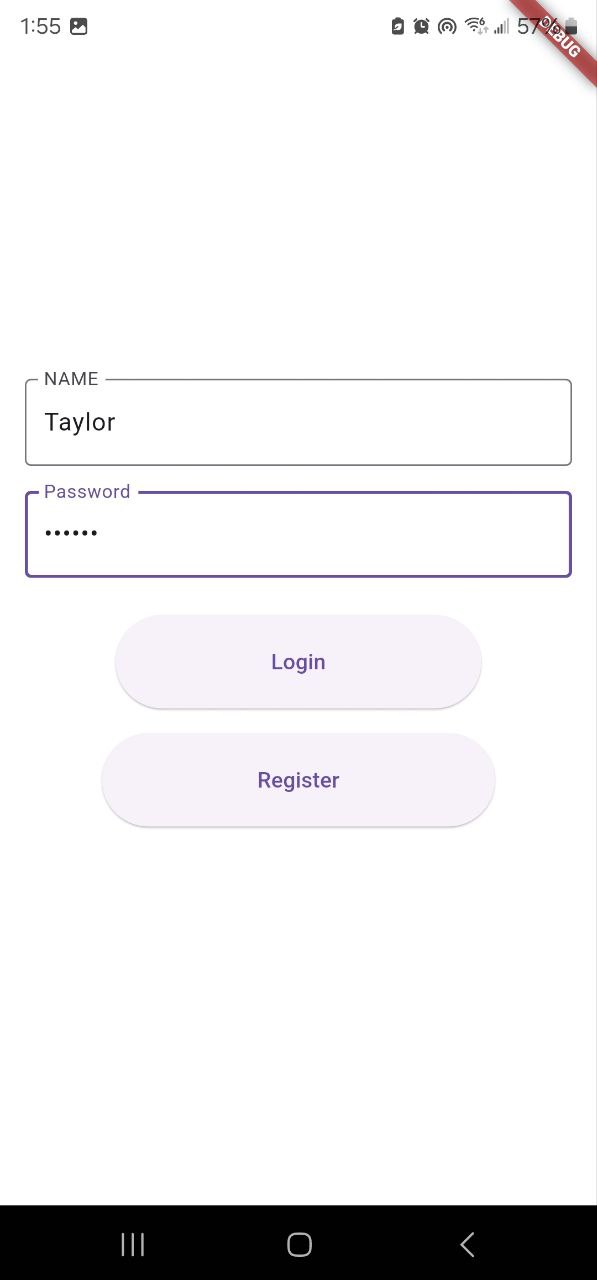}
    \end{minipage}\hfill
    \begin{minipage}{0.2\textwidth}
        \centering
        \includegraphics[width=\linewidth]{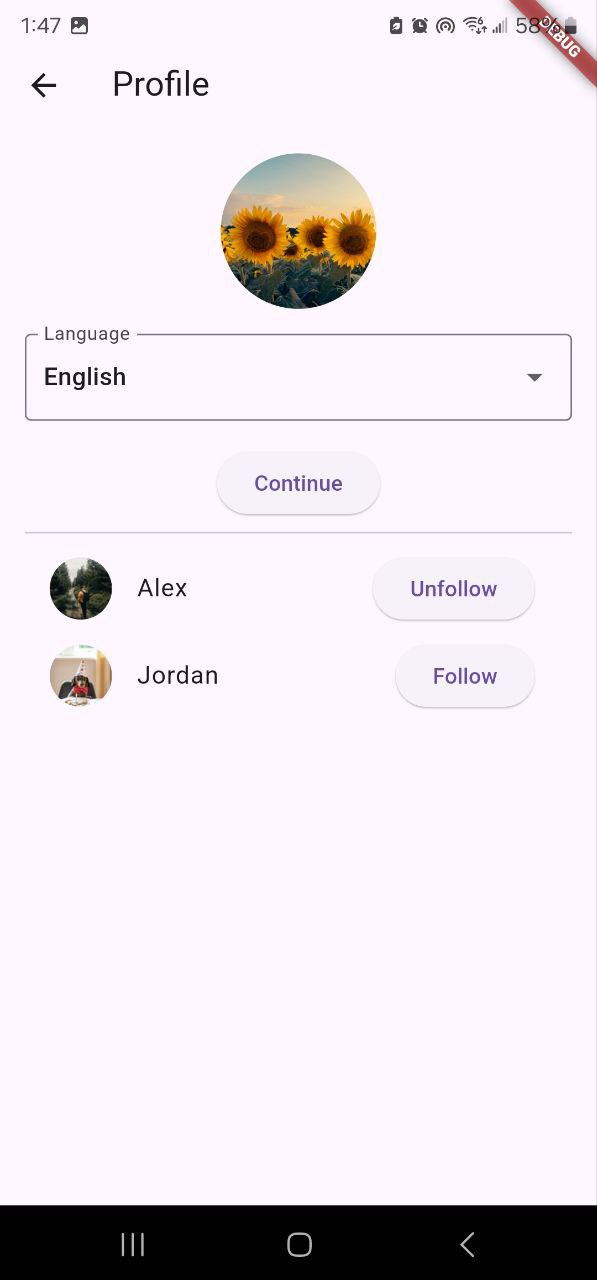}
    \end{minipage}\hspace*{.2in}
    \caption{Overview of the user interface in the application. 
    \textit{Left:} User Login Interface.
    \textit{Right:} User Profile Setup Interface.
    }
    \label{fig:UI}
\end{figure}

The process begins with user registration, where each user creates an account by selecting a username and password Figure~\ref{fig:UI}. After successful registration, users are directed to the profile setup page, where they can upload a profile picture and select a default language. This default language will be used as the primary language for interactions but can be adjusted during subsequent communication, as shown in Figure~\ref{fig:UI}.


User profile metadata, including their chosen language and other personalized settings, is stored securely in a database. Additionally, upon registration, each user is assigned a unique blockchain address that acts as their identifier on the blockchain network. This address is crucial for ensuring the integrity and traceability of their posts and interactions within the system.

\subsection{Timeline and User Interaction}

Once the profile setup is complete, users are redirected to the Timeline page, where they can interact with the posts of users they follow. The timeline is designed to present voice posts from followed users, providing an option to react or interact with the content, as shown in Figure~\ref{fig:UI2}.


A floating action button on the Timeline page allows users to create and submit new posts. When creating a new post, the system pre-selects the user's default language from their profile settings. However, users have the flexibility to change the language for individual posts if necessary.

\subsection{Voice Posting and Automatic Speech Recognition (ASR) Integration}

\begin{figure}[htbp]
    \centering
    \begin{minipage}{0.16\textwidth}
        \centering
        \includegraphics[width=\linewidth]{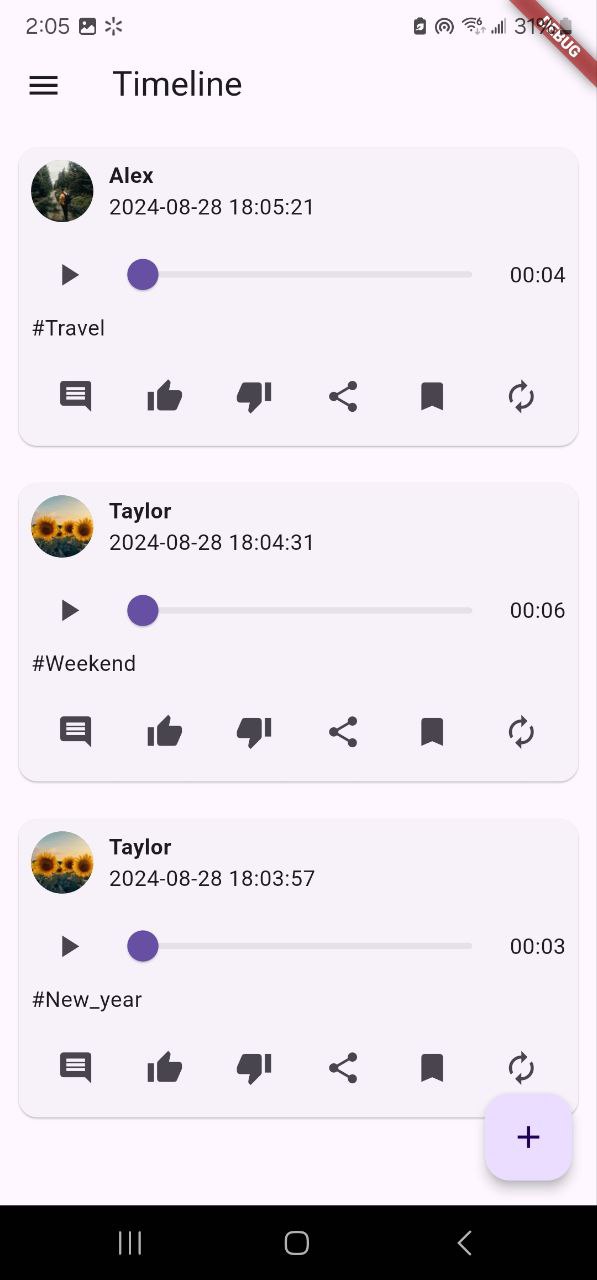}
    \end{minipage}\hfill
    \begin{minipage}{0.16\textwidth}
        \centering
        \includegraphics[width=\linewidth]{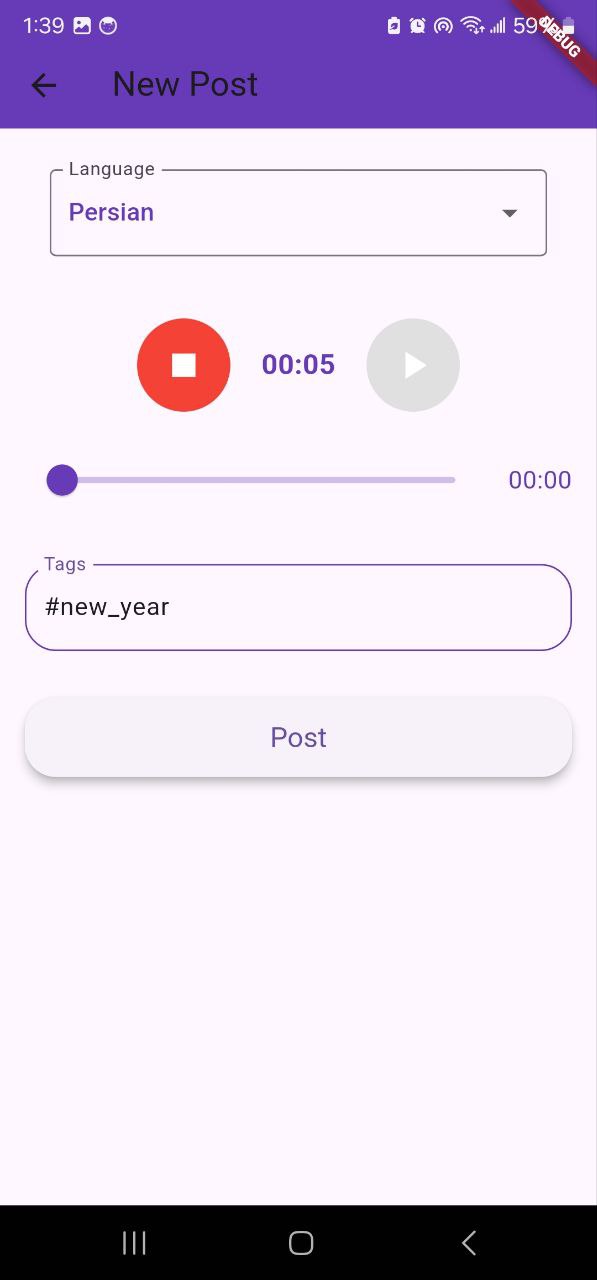}
    \end{minipage}\hfill
    \begin{minipage}{0.16\textwidth}
        \centering
        \includegraphics[width=\linewidth]{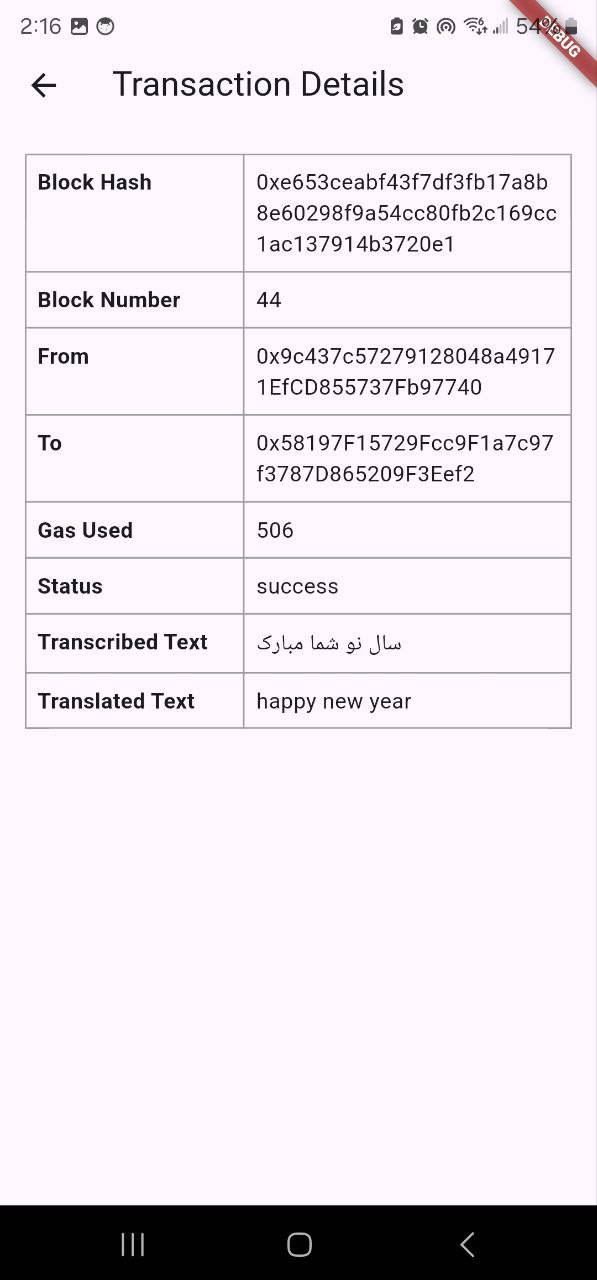}
    \end{minipage}
    \caption{Key interfaces in the application. 
    \textit{Left:} Timeline Interface.
    \textit{Center:} New Post Interface.
    \textit{Right:} Transaction Details Screen.
    }
    \label{fig:UI2}
\end{figure}

To create a post, users record their speech, which is saved as a WAV file. Before submitting, users can replay their recording to ensure accuracy and clarity. Once satisfied, they submit the voice file, which is sent to the local server for processing , as illustrated in Figure~\ref{fig:UI2}.

On the server, the ASR system transcribes the speech to text. The system then calls the blockchain's smart contract API to log the transaction, ensuring that the post's content is recorded immutably on the Ethereum blockchain. This step is essential for verifying the authenticity and integrity of each post.

\subsection{Multilingual Translation and Blockchain Interaction}

Followers of the user who created the post will see the new post on their timeline. If the follower's language matches the original language of the post, they will hear the original voice recording.However, if the languages differ, the system automatically translates the text into the follower's language using AI-based speech-to-text translation (S2TT), and this information is saved in blockchain for transparency and security. Simultaneously, the voice post is translated into the user's language using speech-to-speech translation (S2ST) and played for the user.

All translations and interactions are securely stored on the blockchain, allowing users to verify the sender and the content's integrity through the transaction page, where details like block hash, sender address, and text of the post are displayed by clicking on the transaction button under each post, as shown in Figure~\ref{fig:UI2}.

\section{Evaluation and Results}

The performance of the proposed multilingual voice-based social network was evaluated across several key metrics, as shown in Tables \ref{tab:system_performance}, \ref{tab:user_experience}, and \ref{tab:cross_platform}. Each metric was carefully selected to assess different aspects of the system's performance, ensuring a comprehensive evaluation.

\subsection{System Performance and Blockchain Evaluation}

Table \ref{tab:system_performance} outlines the system's key performance metrics. The ASR system achieved 95.7 percent accuracy in English, while translation accuracy reached 92.3 percent, ensuring reliable communication across languages. The blockchain demonstrated an average transaction time of 1.2 seconds with an overall system latency of 7.8 seconds, confirming the system’s responsiveness. Storage cost per transaction was maintained at 0.0000036 ETH, highlighting its cost-effectiveness.

The ASR Accuracy (English) metric measures the accuracy of the ASR system in converting spoken language into text. With an accuracy of 95.7 percent, the ASR system demonstrates a high level of precision, which is crucial for ensuring that spoken messages are correctly interpreted before being translated. High ASR accuracy reduces errors in subsequent translation and voice synthesis processes, thereby maintaining the integrity of the message as it moves through the system.

Translation Accuracy, recorded at 92.3 percent, assesses the system’s ability to accurately translate text from one language to another. In a multilingual voice-based social network, accurate translations ensure that the original meaning and intent of the message are preserved, which is particularly important in disaster situations where miscommunication can lead to severe consequences.

The Blockchain Transaction Time, averaging 1.2 seconds, measures the speed at which transactions (such as message postings) are processed and recorded on the blockchain. This metric reflects the system's responsiveness, especially in environments where time-sensitive communication is necessary. The choice of blockchain technology provides secure and immutable records of transactions, which is essential for maintaining the authenticity and integrity of the messages.

System Latency (End-to-End), measured at 7.8 seconds, refers to the total time taken from when a user submits a voice message to when the message is fully processed and delivered to the recipient. This metric is crucial in disaster scenarios where timely information exchange can make a significant difference in response efforts.

Storage Cost per Transaction, maintained at 0.0000036 ETH, measures the cost associated with storing each message transaction on the Ethereum blockchain. This metric highlights the cost-effectiveness of the system, particularly in large-scale deployments where numerous transactions need to be recorded securely.

These metrics collectively provide a robust evaluation of the system’s performance, demonstrating its suitability for real-world deployment in disaster communication scenarios.

\begin{table}[h!] 
\centering
\caption{System Performance Metrics and Blockchain Performance Evaluation} 
\begin{tabular}{ll}
\toprule
Metric                       & Value         \\
\midrule
ASR Accuracy (English)                & 95.7 percent                \\
Translation Accuracy                  & 92.3 percent                \\
Blockchain Transaction Time           & 1.2 seconds           \\
System Latency (end-to-end)           & 7.8 seconds           \\
Storage Cost per Transaction          & 0.0000036 ETH         \\
\bottomrule
\end{tabular}
\label{tab:system_performance}
\end{table}

\subsection{User Experience Survey}

Table \ref{tab:user_experience} presents the results of a user experience survey with 11 participants. Key findings include:
\begin{itemize}
    \item Ease of Use: 55 percent rated it excellent.
    \item Translation Accuracy: 46 percent rated it excellent, with room for improvement.
    \item Voice Clarity: 64 percent rated it excellent, reflecting high voice output quality.
    \item Overall Experience: 55 percent rated it excellent, indicating a positive reception.
\end{itemize}

The survey results highlight the system's user-friendliness and overall performance. Ease of Use was rated as excellent by 55 percent of participants, indicating that the interface is intuitive and accessible. Translation Accuracy, though rated excellent by 46 percent, shows some room for improvement, which is expected in a system dealing with multiple languages and cultural nuances. Voice Clarity was highly rated, with 64 percent rating it excellent, underscoring the quality of the voice synthesis component of the system. The Overall Experience rating of 55 percent excellent reflects a generally positive reception among users, pointing to the system’s effectiveness in providing a seamless multilingual communication experience.

\begin{table}[t] 
\centering
\small  
\caption{User Experience Survey Results with 11 Users} 
\begin{tabular}{@{}l@{}c@{\,\,}c@{\,\,}c@{\,\,}c@{}}
\toprule
Category & Excellent (\%) & Good (\%) & Fair (\%) & Poor (\%) \\
\midrule
Ease of Use           & 55 & 36 & 9 & 0 \\
Translation Accuracy  & 46 & 36 & 18 & 0 \\
Voice Clarity         & 64 & 18 & 9 & 9 \\
Overall Experience    & 55 & 27 & 18 & 0 \\
\bottomrule
\end{tabular}
\label{tab:user_experience}
\end{table}

\subsection{Cross-Platform Performance}

The system's performance was evaluated for Android, iOS, and Web platforms using load time, latency, and user ratings to ensure a consistent experience across devices.
Load time measures the duration from the completion of login to when the timeline is fully loaded. The Web platform was the fastest at 1.9 seconds, followed by iOS (2.1 seconds), and Android (2.2 seconds), indicating quick startup across platforms.

Latency refers to the time taken for the system to translate, write to the blockchain, and start playback after a user requests playback. The Web platform showed a latency of 8.6 seconds, Android 7.8 seconds, and iOS 8.7 seconds, demonstrating the system's responsiveness in processing and delivering content.

User ratings reflect overall satisfaction with the application on a scale of one to five, where one represents poor performance and five represents excellent performance. This scale is commonly used because it provides a straightforward, intuitive measure of user experience that is easy for participants to understand and respond to. Ratings were collected through surveys where users rated their experience based on factors such as ease of use, responsiveness, and overall satisfaction. The average rating for each platform was calculated by taking the mean of all ratings provided by users. Android received the highest average rating of 4.5, followed by Web at 4.3, and iOS at 4.1. These ratings indicate that the application meets user expectations and provides a generally positive experience across platforms.

These results confirm that the system is well-optimized for various platforms, ensuring both accessibility and reliability.

\begin{table}[h!] 
\centering
\caption{Cross-Platform Performance Comparison} 
\begin{tabular}{lccc}
\toprule
Platform & Load Time (s) & Latency (s) & Rating (1-5) \\
\midrule
Android  & 2.2 & 7.8 & 4.5 \\
iOS      & 2.1 & 8.7 & 4.1 \\
Web      & 1.9 & 8.6 & 4.3 \\
\bottomrule
\end{tabular}
\label{tab:cross_platform}
\end{table}

Overall, the system provides effective, secure, and user-friendly communication across multiple platforms, supported by reliable blockchain integration.

\section{Discussion}

The integration of Ethereum blockchain with AI-driven multilingual translation has proven to be an effective solution for secure and reliable communication, particularly in high-stakes scenarios like disaster situations. The system requirements for running this application effectively include a minimum GPU of 8 GB, an operating system of Ubuntu or macOS, 16 GB of RAM, and 100 GB of SSD storage.
Our implementation leverages several key technologies to achieve these goals:


For the blockchain component, we utilized Ganache, a personal Ethereum blockchain designed for development and testing. Ganache provided a local blockchain instance that enabled us to deploy smart contracts in a controlled environment, simulating the conditions of a live network. This allowed for thorough testing of our smart contracts and their interactions with the application, ensuring robustness before any potential deployment to a production environment.

Truffle was employed as the development framework, offering a suite of tools for compiling and deploying smart contracts to the Ganache instance. Truffle's asset pipeline streamlined the deployment process, allowing us to write deployment scripts that automatically pushed our contracts to the blockchain. The pre-configured accounts in Ganache, complete with Ether for testing, simplified the validation of contract functionality and transaction handling.


The Meta SeamlessM4T model was central to our AI speech processing, enabling highly accurate ASR and real-time multilingual translation. To balance accuracy and processing speed, we opted for the medium-sized model, which offered an optimal trade-off between the two, ensuring that the system remained responsive while still delivering precise translations. This AI model facilitated the seamless conversion of speech to text and then into multiple languages, maintaining the original intent and context across translations. The model covers 37 languages for S2ST, including Modern Standard Arabic, Bengali, Catalan, Czech, Mandarin Chinese, Welsh, Danish, German, English, Estonian, Finnish, French, Hindi, Indonesian, Italian, Japanese, Korean, Maltese, Dutch, Persian, Polish, Portuguese, Romanian, Russian, Slovak, Spanish, Swedish, Swahili, Telugu, Tagalog, Thai, Turkish, Ukrainian, Urdu, Northern Uzbek, and Vietnamese. The model’s efficiency and accuracy were particularly beneficial in a disaster context, where timely and precise communication is critical.


Flutter was chosen for the development of our application due to its capability to create cross-platform solutions with a single codebase. This choice ensured that our application was accessible on multiple devices and operating systems, ranging from mobile phones to desktops. Flutter's flexibility allowed us to maintain a consistent user experience across all platforms, which is crucial in disaster scenarios where users might have access to different types of devices.

We utilized PostgreSQL as the database system for managing essential data, chosen for its robustness, scalability, and efficiency in handling complex queries. It stored user profile information (e.g., usernames, language preferences), following data, and post metadata (e.g., tags, languages). This setup enabled personalized user experiences, efficient content filtering, and secure multilingual functionality, complementing our AI-driven translation and blockchain technologies.




\section{Conclusion}

In this paper, we have presented a novel approach to multilingual communication in disaster scenarios through the development of a voice-based social network that integrates advanced AI and blockchain technologies. The application we developed leverages Meta’s SeamlessM4T generative AI to provide real-time, accurate translations of voice messages, enabling seamless communication across language barriers. By processing all data locally and utilizing Ethereum blockchain for secure message storage, our solution ensures both the reliability and integrity of communication, which is crucial in high-stakes environments where external systems might fail.

Our system’s cross-platform compatibility, facilitated by the use of Flutter, ensures accessibility across a wide range of devices, making it versatile and practical in diverse disaster situations. The integration of PostgreSQL for data management further enhances the application’s ability to store and retrieve user information efficiently, providing a personalized user experience while maintaining high security standards.

Implementing real-time sentiment analysis could allow the system to automatically detect distress or urgency in user communications, triggering emergency alerts to responders and potentially accelerating aid. Additionally, using text processing for auto-tagging could categorize and prioritize messages based on their content, further optimizing the response process. Moreover, integrating Meta’s SeamlessExpressive model could be another promising avenue for future development. This model aims to preserve the expression and intricacies of speech across languages, ensuring that the emotional tone and subtleties of communication are maintained even after translation. This enhancement could significantly improve the effectiveness of communication during disaster scenarios, where the emotional context of messages is often as important as the content itself. Additionally, using text processing for auto-tagging could categorize and prioritize messages based on their content, further optimizing the response process.

Moreover, integrating physiological data with text processing, as explored in recent research, could enhance the tagging process by adding contextual awareness, leading to more accurate and responsive communication in critical situation~\cite{dongre2024integrating}.

By addressing these areas, the application can be further refined to provide not only accurate and secure communication but also emotionally resonant and contextually aware interactions, making it an even more powerful tool in disaster response and management.

\bibliographystyle{unsrt}  

\end{document}